\documentclass{webofc}
\usepackage[varg]{txfonts}   
\begin{document}

\title{Study of hyperon-nucleon interactions at BESIII}

\author{\firstname{Jielei} \lastname{Zhang}\inst{1}\fnsep\thanks{\email{zhangjielei@henu.edu.cn}}\fnsep\thanks{This work is supported by National Natural Science Foundation of China (NSFC) under Contracts No. 12375071.} on behalf of the BESIII Collaboration
}

\institute{School of Physics and Electronics, Henan University, Kaifeng 475004, People's Republic of China
}

\abstract{
  Hyperon-nucleon interactions are important to understand quantum chromodynamics and so-called ``hyperon puzzle'' of neutron star, but limited by the availability and short-lifetime of hyperon beams, the progress of relevant research is very slow. A novel method is used to study hyperon-nucleon interactions based on hyperons produced in the decays of 10 billion $J/\psi$ events collected with the BESIII detector at the BEPCII storage ring, and the target material is beam pipe. The reactions $\Xi^{0}n\rightarrow\Xi^{-}p$ and $\Lambda N\rightarrow\Sigma^+X$ have been observed and measured at BESIII. This is the first study of hyperon-nucleon interactions in electron-positron collisions and opens up a new direction for such research.
}

\maketitle

\section{Introduction}
\label{intro}
Scattering experiments of high energy particle beams bombarding target materials have been of great significance for studying the inner structure of matter and the fundamental interactions. More than 100 years ago, Rutherford bombarded a gold foil with $\alpha$ particles, leading to the proposal of the atomic model~\cite{intro1}, which opened the door for understanding the inner structure of atoms. Later, protons and neutrons as the subconstituents of the atomic nucleus were discovered~\cite{intro2, intro3}, likewise. Furthermore, different kinds of particle beams, such as $\pi^{\pm}$, $K^{\pm}$, and $p/\bar{p}$ beams, led to a series of breakthrough discoveries, like the first observation of the $J/\psi$ charmonium and excited baryons~\cite{intro4, intro5}. Charged long-lived particle beams such as $\pi^{\pm}/K^{\pm}$ can be easily produced in experiments. However, due to significant shorter lifetimes and higher masses, particle beams of hyperons, such as $\Lambda/\Sigma/\Xi$, are more difficult to produce experimentally and corresponding experiments are rare~\cite{intro5}, although measurements of these beams bombarding target material are crucial for understanding quantum chromodynamics.

The experimental study on the interaction between hyperons and target materials began in the 1960s and has lasted for more than half a century~\cite{intro5}. The experimental methods are generally to bombard a hydrogen bubble chamber or a scintillating fiber target with a $K^-$ meson beam to obtain the required hyperons, which allows access to the study of hyperon-nucleon interactions. However, the intensity of hyperon beams produced by these experiments are relatively low and relevant experimental measurements are very scarce. After stagnating for decades, in 2021 and 2022, the CLAS and J-PARC E40 Collaborations reported the latest results of $\Lambda/\Sigma$ and nucleon interactions respectively~\cite{intro6, intro7, intro8, intro9}, which have enhanced the understanding of hyperon-nucleon interactions further. The hyperon-nucleon interactions have been studied by some theoretical models, including the constituent quark model~\cite{intro10, intro11, intro12}, the meson-exchange picture~\cite{intro13, intro14}, and the chiral effective field theory approach~\cite{intro15, intro16, intro17, intro18, intro19}. More experimental measurements are strongly needed to constrain the theoretical models, which can greatly promote the research in this field. Furthermore, the study of hyperon-nucleon interactions is also important to understand the role of hyperons in dense neutron-star matter, to determine the equation of state (EoS) of nuclear matter at supersaturation densities and to understand the so-called ``hyperon puzzle" of neutron stars~\cite{intro20, intro21, intro22}.

\begin{figure}[h]
\centering
\includegraphics[width=5cm,clip]{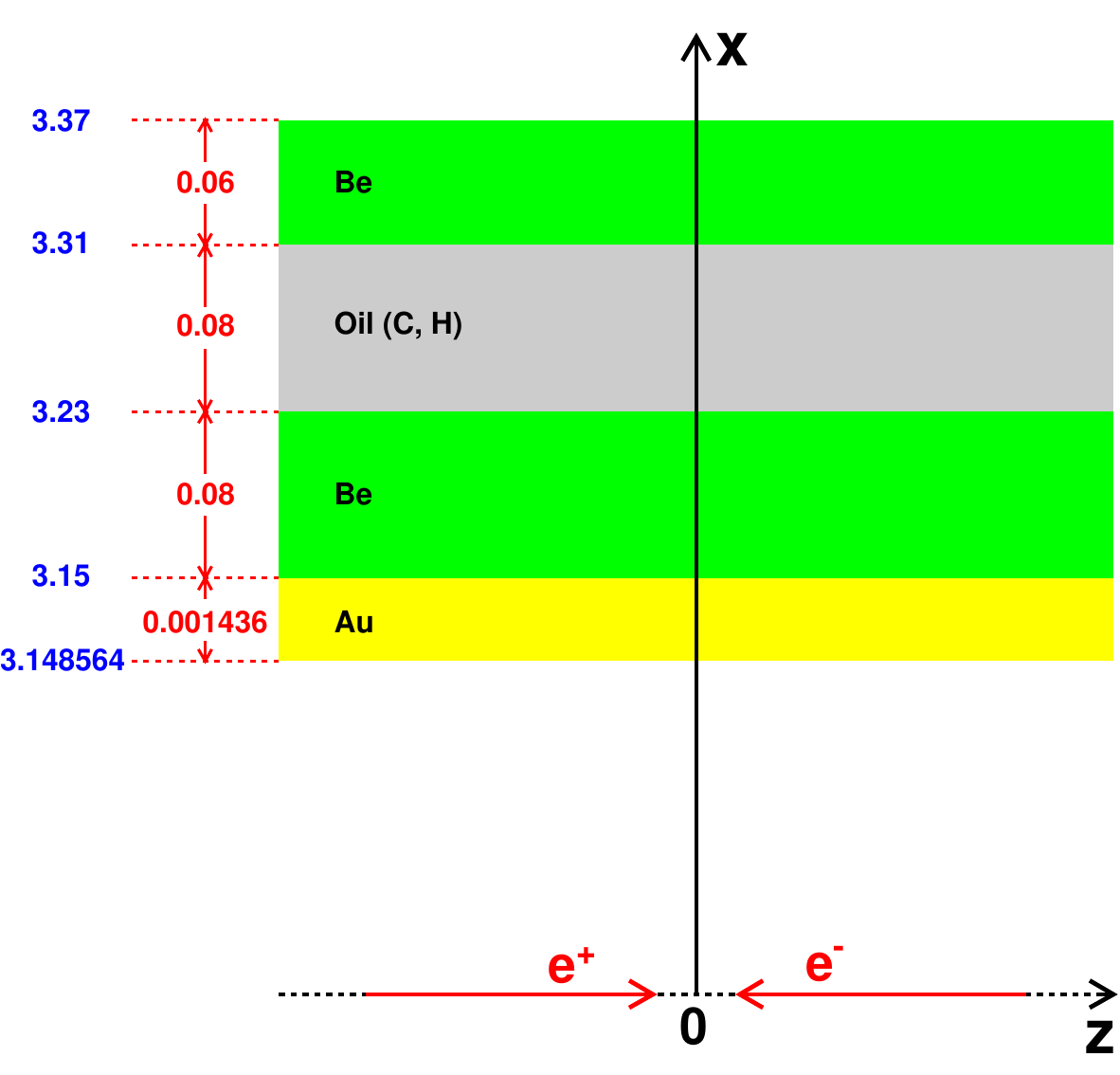}
\caption{Schematic diagram of the beam pipe, the length units are centimeter (cm). The $z$-axis is the symmetry axis of the MDC, and the $x$-axis is perpendicular to the $e^+e^-$ beam direction.}
\label{fig-1}
\end{figure}

The BESIII detector records symmetric $e^+e^-$ collisions at the BEPCII collider~\cite{bepcii}. Details of the BESIII detector can be found in Ref.~\cite{besiii}. With a sample of $(1.0087\pm0.0044)\times10^{10}$ $J/\psi$ events collected by the BESIII detector~\cite{totalnumber}, substantial hyperons can be produced in the decays of $J/\psi$. The hyperons can interact with the material in the beam pipe adjacent to the $e^+e^-$ beam, providing a novel source to study the hyperon-nucleon interactions~\cite{intro23, intro24}. The material of the beam pipe is composed of gold ($^{197}\rm{Au}$), beryllium ($^{9}\rm{Be}$) and oil $(^{12}\rm{C}:$$^{1}\rm{H}$$=1:2.13)$, as shown in Fig.~\ref{fig-1}. In this proceeding, the results related to the reactions $\Xi^{0}n\rightarrow\Xi^{-}p$ and $\Lambda N\rightarrow\Sigma^+X$ are reported~\cite{ref1, ref2}, where $N$ and $X$ represent nucleus and anything, respectively.

\section{First study of reaction $\Xi^{0}n\rightarrow\Xi^{-}p$ at an electron-positron collider}
\label{sec-1}

\begin{figure}[h]
\centering
\includegraphics[width=6cm,clip]{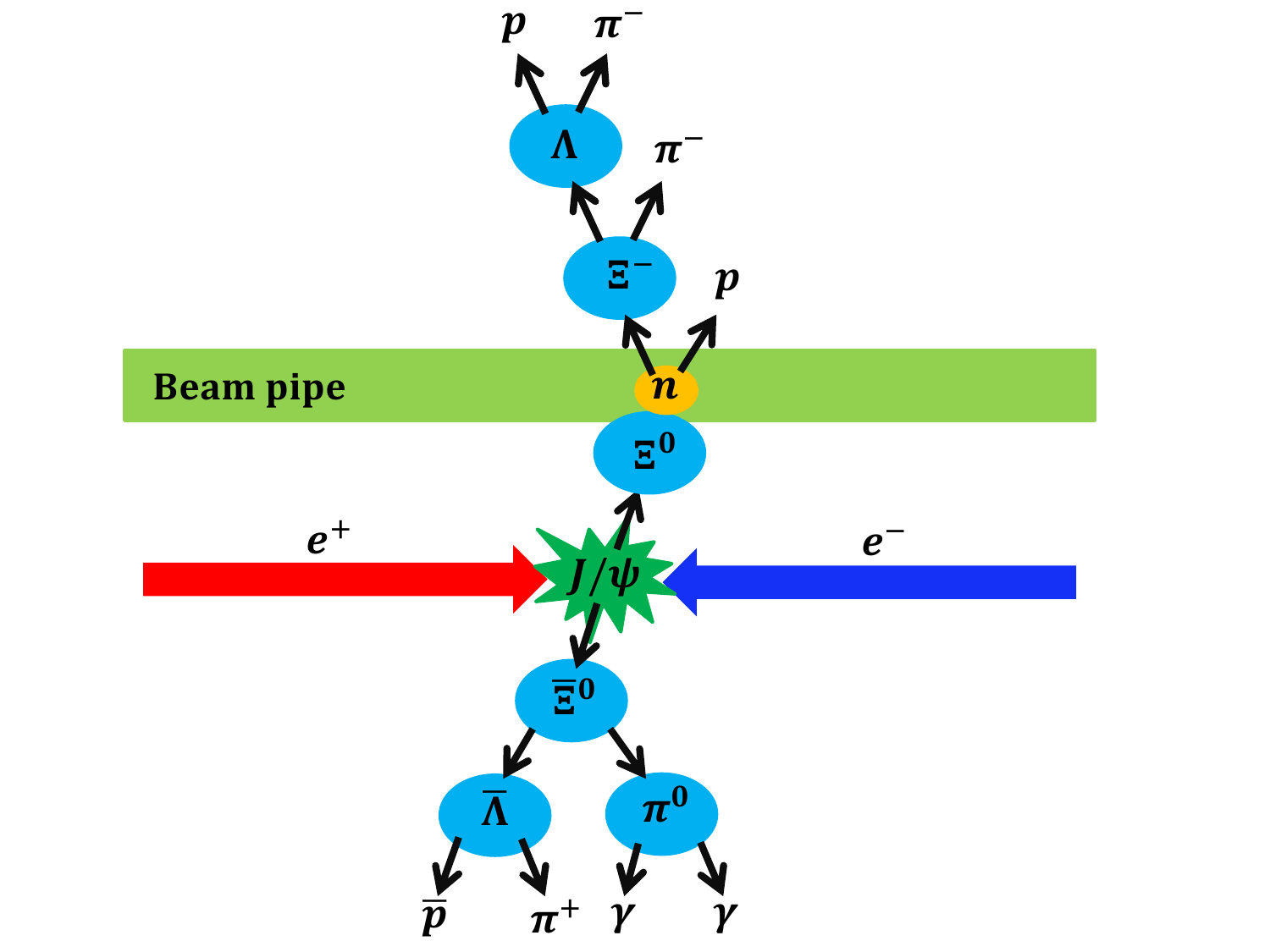}
\caption{A complete topology diagram of the signal process $\Xi^{0}n\rightarrow\Xi^{-}p$.}
\label{fig-2}
\end{figure}

Using $(1.0087\pm0.0044)\times10^{10}$ $J/\psi$ events collected with the BESIII detector at the BEPCII storage ring, the process $\Xi^{0}n\rightarrow\Xi^{-}p$ is studied, where the $\Xi^0$ baryon is produced in the process $J/\psi\rightarrow\Xi^0\bar{\Xi}^0$ and the neutron is a component of the $^9\rm{Be}$, $^{12}\rm{C}$ and $^{197}\rm{Au}$ nuclei in the beam pipe. The signal process considered in this work is $J/\psi\rightarrow\Xi^0\bar{\Xi}^0$, $\Xi^{0}n\rightarrow\Xi^{-}p$, $\Xi^-\rightarrow\Lambda\pi^-$, $\Lambda\rightarrow p\pi^-$, $\bar{\Xi}^0\rightarrow\bar{\Lambda}\pi^0$, $\bar{\Lambda}\rightarrow\bar{p}\pi^+$, $\pi^0\rightarrow\gamma\gamma$, as shown in Fig.~\ref{fig-2}. The analysis method is that we use $\bar{\Xi}^0$ to tag the event and require the recoiling mass to be in the $\Xi^{0}$ mass region, then reconstruct $\Xi^-$ and $p$ in the signal side.

\begin{figure}[h]
\centering
\includegraphics[width=5cm,clip]{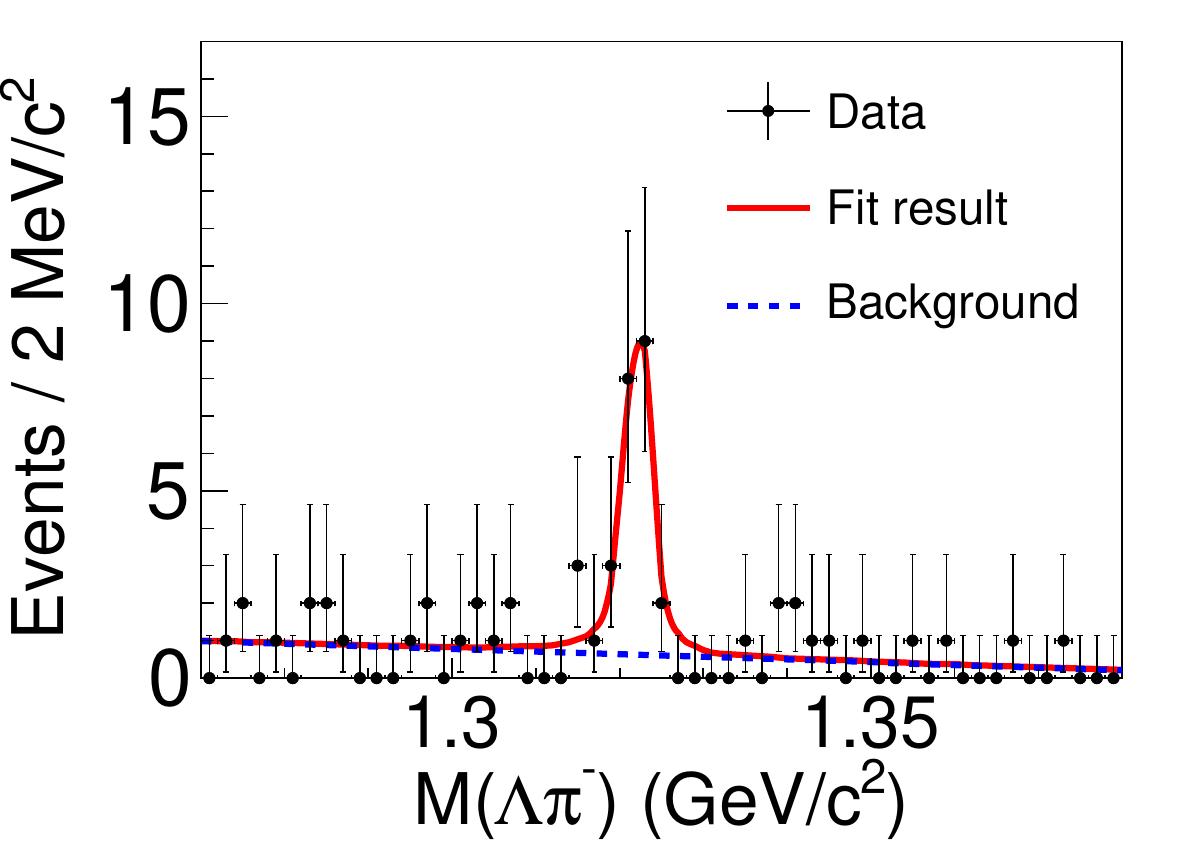}
\caption{Distribution of $M(\Lambda\pi^-)$ in data (dots with error bars). The red solid curve is the fit result, which includes signal component and background component. The blue dashed curve is the background component.}
\label{fig-3}
\end{figure}

Figure~\ref{fig-3} shows the $M(\Lambda\pi^-)$ distribution from data after final event selection, a clear $\Xi^-$ signal is observed with a statistical significance of $7.1\sigma$, corresponding to the reaction $\Xi^{0}n\rightarrow\Xi^{-}p$. The cross section of the reaction $\Xi^0+{^9\rm{Be}}\rightarrow\Xi^-+p+{^8\rm{Be}}$ is determined to be $\sigma(\Xi^0+{^9\rm{Be}}\rightarrow\Xi^-+p+{^8\rm{Be}})=(22.1\pm5.3_{\rm{stat}}\pm4.5_{\rm{sys}})$~mb at the $\Xi^0$ momentum of about $0.818$~GeV/$c$. If the effective number of reaction neutrons in a $^9\rm{Be}$ nucleus is taken as $3$~\cite{intro25}, the cross section of $\Xi^{0}n\rightarrow\Xi^{-}p$ for a single neutron is determined to be $\sigma(\Xi^{0}n\rightarrow\Xi^{-}p)=(7.4\pm1.8_{\rm{stat}}\pm1.5_{\rm{sys}})$~mb, consistent with theoretical predictions in Refs.~\cite{intro12, intro16, intro18}. Furthermore, we do not observe any significant $H$-dibaryon signal in the $\Xi^-p$ final state for this reaction process~\cite{dibaryon1, dibaryon2}. This work is the first study of hyperon-nucleon interactions in electron-positron collisions, and opens up a new direction for such research.

\section{First measurement of $\Lambda N\rightarrow\Sigma^+X$ with $\Lambda$ from $e^+e^-\rightarrow J/\psi\rightarrow\Lambda\bar{\Lambda}$}
\label{sec-2}

Using an $e^+e^-$ collision data sample of $(1.0087\pm0.0044)\times10^{10}$ $J/\psi$ events taken at BESIII, the process $\Lambda N\rightarrow\Sigma^+X$ is studied for the first time. The $\Sigma^+$ hyperons are produced by the collisions of $\Lambda$ hyperons from $J/\psi$ decays with nuclei in the material of the BESIII detector. The signal process in this work is $J/\psi\rightarrow\Lambda\bar{\Lambda}$, $\Lambda N\rightarrow\Sigma^+X$, $\Sigma^+\rightarrow p\pi^0$, $\pi^0\rightarrow\gamma\gamma$, $\bar{\Lambda}\rightarrow\bar{p}\pi^+$. The analysis method is that we use $\bar{\Lambda}$ to tag the event and require the recoiling mass to be in the $\Lambda$ mass region, then only reconstruct $\Sigma^+$ in the signal side.

\begin{figure}[h]
\centering
\includegraphics[width=6cm,clip]{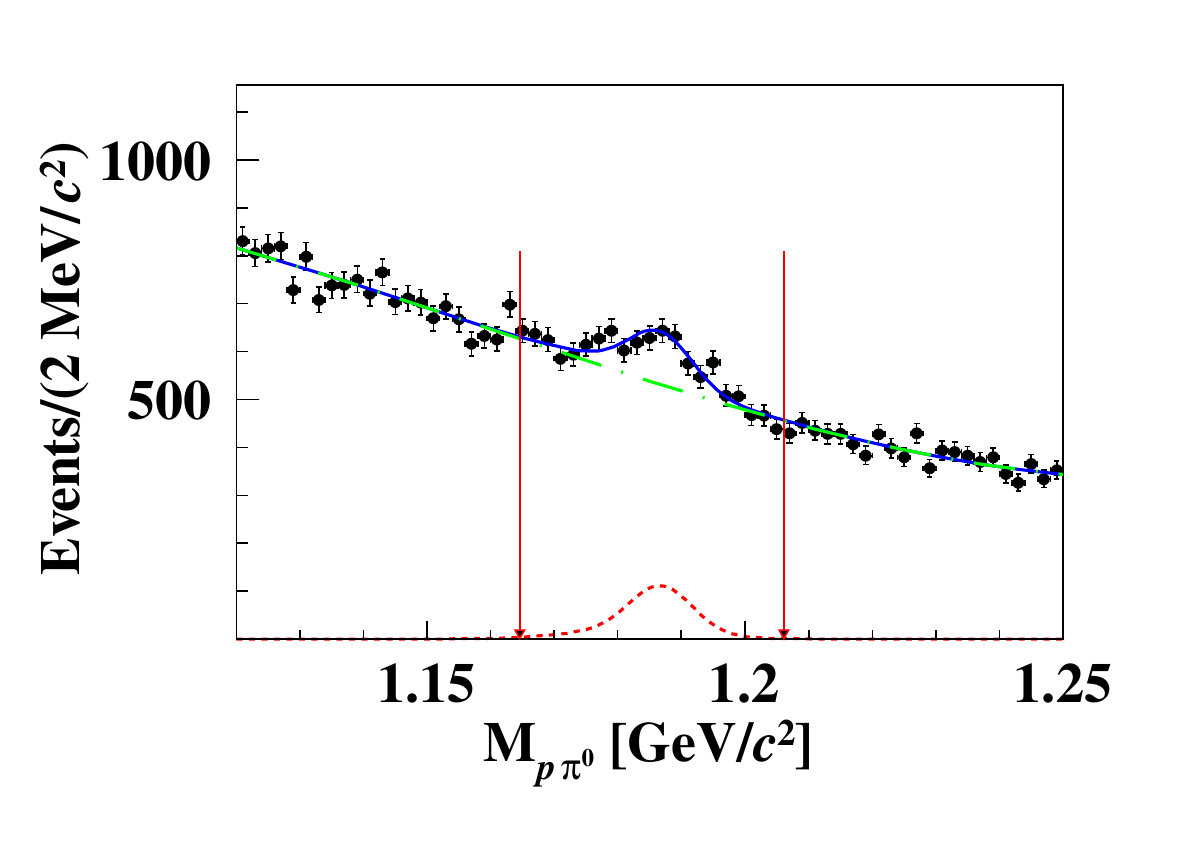}
\caption{The $M_{p\pi^0}$ distribution with the fit result overlaid. The black dots with error bars represent the data. The blue solid line is the fit result, which includes signal component and background component. The dashed red line is the signal and the dot-dashed green line is the background. The red arrows indicate the signal range.}
\label{fig-4}
\end{figure}

Figure~\ref{fig-4} shows the $M_{p\pi^0}$ distribution where a clear $\Sigma^+$ signal is observed, corresponding to the reaction $\Lambda N\rightarrow\Sigma^+X$. The total cross section of $\Lambda+{^{9}\rm{Be}}\rightarrow\Sigma^++X$ is measured to be $\sigma(\Lambda+{^{9}\rm{Be}}\rightarrow\Sigma^++X)=(37.3\pm4.7_{\rm{stat}}\pm3.5_{\rm{sys}})$~mb at the $\Lambda$ momentum of about $1.074$~GeV/$c$. Taking 1.93 as the ratio of the cross section of $\Lambda+{^{9}\rm{Be}}\rightarrow\Sigma^++X$ and $\Lambda p\rightarrow\Sigma^+X$ by assuming the signal process as a surface reaction, the cross section of $\Lambda p\rightarrow\Sigma^+X$ is determined to be $\sigma(\Lambda p\rightarrow\Sigma^+X)=(19.3\pm2.4_{\rm{stat}}\pm1.8_{\rm{sys}})$~mb. By virtue of charge independence, the cross section of $\Lambda p\rightarrow\Sigma^+n$ is just twice that of $\Lambda p\rightarrow\Sigma^0p$~\cite{intro26}, the measured result is consistent with previous experiments regarding the cross section measurements of $\Lambda p\rightarrow\Sigma^0p$~\cite{intro27}. This analysis is the first study of $\Lambda$-nucleon interactions at an $e^+e^-$ collider, providing useful information for understanding baryon-baryon interactions.

\section{Summary and outlook}
\label{summary}

In summary, the hyperon-nucleon interactions are studied using a novel method at BESIII, the hyperons are produced in the decays of $J/\psi$ events and the beam pipe can be taken as the target material. The reactions $\Xi^{0}n\rightarrow\Xi^{-}p$ and $\Lambda N\rightarrow\Sigma^+X$ are observed and measured at BESIII, which can enhance the understanding of hyperon-nucleon interactions. This is the first study of hyperon-nucleon interactions in electron-positron collisions and opens up a new direction for such research.

Using the same method, other hyperon-nucleon reactions can be studied at BESIII. Especially, experimental information on antihyperon-nucleon interactions can also be provided, so more interesting results will come out soon. Furthermore, we may be able to design targets of specific materials to study hyperon-nucleon interactions in future super tau-charm facilities~\cite{super1, super2}. With more statistics at that time, we can also study the momentum-dependent cross section or differential cross section distributions based on the hyperons from multi-body decays of $J/\psi$ or other charmonia.


\begin{thebibliography}{}

\bibitem{intro1} E.~Rutherford, Phil. Mag. Ser. 6 \textbf{21}, 669 (1911).

\bibitem{intro2} E.~Rutherford, Phil. Mag. Ser. 6 \textbf{37}, 581 (1919).

\bibitem{intro3} J.~Chadwick, Nature \textbf{129}, 312 (1932).

\bibitem{intro4} J.~J.~Aubert \textit{et al.} [E598 Collaboration], Phys. Rev. Lett. \textbf{33}, 1404 (1974).

\bibitem{intro5} R.~L.~Workman \textit{et al.} [Particle Data Group], PTEP \textbf{2022}, 083C01 (2022).

\bibitem{intro6} J.~Rowley \textit{et al.} [CLAS Collaboration], Phys. Rev. Lett. \textbf{127}, 272303 (2021).

\bibitem{intro7} K.~Miwa \textit{et al.} [J-PARC E40 Collaboration], Phys. Rev. C \textbf{104}, 045204 (2021).

\bibitem{intro8} T.~Nanamura \textit{et al.} [J-PARC E40 Collaboration], PTEP \textbf{2022}, 093D01 (2022).

\bibitem{intro9} K.~Miwa \textit{et al.} [J-PARC E40 Collaboration], Phys. Rev. Lett. \textbf{128}, 072501 (2022).

\bibitem{intro10} C.~Nakamoto, Y.~Fujiwara and Y.~Suzuki, Nucl. Phys. A \textbf{639}, 51 (1998).

\bibitem{intro11} V.~G.~J.~Stoks and T.~A.~Rijken, Phys. Rev. C \textbf{59}, 3009 (1999).

\bibitem{intro12} Y.~Fujiwara, M.~Kohno, C.~Nakamoto and Y.~Suzuki, Phys. Rev. C \textbf{64}, 054001 (2001).

\bibitem{intro13} M.~Yamaguchi, K.~Tominaga, T.~Ueda and Y.~Yamamoto, Prog. Theor. Phys. \textbf{105}, 627 (2001).

\bibitem{intro14} J.~Haidenbauer and U.~G.~Mei\ss{}ner, Phys. Rev. C \textbf{72}, 044005 (2005).

\bibitem{intro15} H.~Polinder, J.~Haidenbauer and U.~G.~Meissner, Phys. Lett. B \textbf{653}, 29 (2007).

\bibitem{intro16} J.~Haidenbauer, U.~G.~Mei\ss{}ner and S.~Petschauer, Nucl. Phys. A \textbf{954}, 273 (2016).

\bibitem{intro17} K.~W.~Li, X.~L.~Ren, L.~S.~Geng and B.~W.~Long, Chin. Phys. C \textbf{42}, 014105 (2018).

\bibitem{intro18} J.~Haidenbauer and U.~G.~Mei\ss{}ner, Eur. Phys. J. A \textbf{55}, 23 (2019).

\bibitem{intro19} J.~Song, Z.~W.~Liu, K.~W.~Li and L.~S.~Geng, Phys. Rev. C \textbf{105}, 035203 (2022).

\bibitem{intro20} D.~Chatterjee and I.~Vida\~na, Eur. Phys. J. A \textbf{52}, 29 (2016).

\bibitem{intro21} I.~Vida\~na, Proc. Roy. Soc. Lond. A \textbf{474}, 0145 (2018).

\bibitem{intro22} L.~Tolos and L.~Fabbietti, Prog. Part. Nucl. Phys. \textbf{112}, 103770 (2020).

\bibitem{bepcii} C.~H.~Yu \textit{et al.}, Proceedings of IPAC2016, Busan, Korea, 2016, doi:10.18429/JACoW-IPAC2016-TUYA01.

\bibitem{besiii} M.~Ablikim \textit{et al.} [BESIII Collaboration], Nucl. Instrum. Meth. A \textbf{614}, 345 (2010).

\bibitem{totalnumber} M.~Ablikim \textit{et al.} [BESIII Collaboration], Chin. Phys. C \textbf{46}, 074001 (2022).

\bibitem{intro23} C.~Z.~Yuan and M.~Karliner, Phys. Rev. Lett. \textbf{127}, 012003 (2021).

\bibitem{intro24} J.~P.~Dai, H.~B.~Li, H.~Miao and J.~Y.~Zhang, arXiv:2209.12601.

\bibitem{ref1} M.~Ablikim \textit{et al.} [BESIII Collaboration], Phys. Rev. Lett. \textbf{130}, 251902 (2023).

\bibitem{ref2} M.~Ablikim \textit{et al.} [BESIII Collaboration], arXiv: 2310.00720.

\bibitem{intro25} J.~K.~Ahn, S.~Aoki, K.~S.~Chung, M.~S.~Chung, H.~En'yo, T.~Fukuda, H.~Funahashi, Y.~Goto, A.~Higashi and M.~Ieiri, \textit{et al.} Phys. Lett. B \textbf{633}, 214 (2006).

\bibitem{dibaryon1} R.~L.~Jaffe, Phys. Rev. Lett. \textbf{38}, 195 (1977).

\bibitem{dibaryon2} A.~T.~M.~Aerts, P.~J.~G.~Mulders and J.~J.~de Swart, Phys. Rev. D \textbf{17}, 260 (1978).

\bibitem{intro26} J.~A.~Kadyk, G.~Alexander, J.~H.~Chan, P.~Gaposchkin and G.~H.~Trilling, Nucl. Phys. B \textbf{27}, 13 (1971).

\bibitem{intro27} J.~M.~Hauptman, J.~A.~Kadyk and G.~H.~Trilling, Nucl. Phys. B \textbf{125}, 29 (1977).

\bibitem{super1} A.~E.~Bondar \textit{et al.} [Charm-Tau Factory Collaboration], Phys. Atom. Nucl. \textbf{76}, 1072 (2013).

\bibitem{super2} M.~Achasov \textit{et al.}, Front. Phys. \textbf{19}, 14701 (2024).

\end{thebibliography}
\end{document}